\documentclass[pop,aip,preprint,amsmath,amssymb]{revtex4-1}

\usepackage{graphicx}% Include figure files
\usepackage{bm}% bold math
\renewcommand{\vec}[1]{\boldsymbol{#1}}

\def\aap{{\itshape Astron. Astrophys.} }
\def\apj{{\itshape Astrophys. J.} }
\def\jgr{{\itshape J. Geophys. Res.} }
\def\jpp{{\itshape J. Plasma Phys.} }
\def\mnras{{\itshape Mon. Not. R. Astron. Soc.} }
\def\pop{{\itshape Phys. Plasmas} }
\def\prl{{\itshape Phys. Rev. Lett.} }

\begin{document}

\title[PHYSICS OF PLASMAS]{Magnetohydrodynamic structure of a plasmoid
in fast reconnection in low-beta plasmas: Shock-shock interactions}

\author{Seiji Zenitani}
\affiliation{National Astronomical Observatory of Japan, 2-21-1 Osawa, Mitaka, Tokyo 181-8588, Japan.}
\email{seiji.zenitani@nao.ac.jp}

\date{Submitted 17 January 2015; accepted 5 March 2015}

\begin{abstract}
The shock structure of a plasmoid in magnetic reconnection in low-beta plasmas
is investigated by two-dimensional magnetohydrodynamic simulations.
Using a high-accuracy code with unprecedented resolution,
shocks, discontinuities, and their intersections are resolved and clarified.
Contact discontinuities emanate from triple-shock intersection points,
separating fluids of different origins.
Shock-diamonds inside the plasmoid appear to decelerate a supersonic flow.
New shock-diamonds and a slow expansion fan are found
inside the Petschek outflow.
A sufficient condition for the new shock-diamonds and
the relevance to astrophysical jets are discussed.
\end{abstract}

\maketitle

\section{Introduction}
Magnetic reconnection provides fast plasma transport
in solar, space, laboratory, and astrophysical plasmas. 
By breaking a magnetic topology,
it releases the stored magnetic energy into the energy of a fast outflow jet,
whose speed is approximated by the Alfv\'{e}n speed in the upstream region. 
The reconnection system often involves a ``plasmoid,''
a dense plasma island confined in a magnetic loop.
For example,
when the reconnection jet sweeps plasmas and the reconnected magnetic fields,
a plasmoid is formed at the jet front.
For example,
the tearing-type instabilities generate plasmoids
in a stationary or reconnecting current sheet.

In magnetotail physics,
motivated by satellite observations,\citep{hones77,slavin84}
the structure and dynamics of a plasmoid have long been studied
since the early days of magnetohydrodynamic (MHD) simulations.
The plasmoid has been considered as an outcome of Petschek-type fast reconnection.
Carrying out a series of MHD simulations,
\citet{ugai95} extensively studied the dynamics of a plasmoid
in the downstream of the reconnection outflow.
It was found that
the plasmoid moves at some fraction of the Alfv\'{e}n speed and
that it exhibits a complicated structure, surrounded by slow-shocks.
\citet{shuei01} carried out a large-scale MHD simulation and
they found a pair of intermediate shocks inside the plasmoid,
where the plasmoid swallows the thin plasma sheet. 
Recently, \citet{zeni11a} (hereafter referred to as Paper I)
have introduced a shock-capturing code to the numerical research on the plasmoid structure. 
They found a new class of shocks near the plasmoid.
The shocks are attributed to the low-beta condition of $\beta < 1$,
because the typical speed of the plasmoid motion,
the Alfv\'{e}n speed ($c_{A}$), exceeds the local sound speed ($c_s$).
These new features suggest that
plasmoid physics may be qualitatively different in this regime.
Since the plasma beta is one of the basic parameters in MHD,
plasmoid physics in low-beta plasmas deserves further investigation.

Plasmoids draw recent attention in broader context of MHD reconnection.
For example, \citet{loureiro07} predicted that
a sufficiently long Sweet--Parker (SP) current sheet is
unstable to a plasmoid formation.
As a result, the current sheet is filled with many plasmoids,
and then the SP reconnection switches to
the ``plasmoid-dominated reconnection''
\citep{bhattacharjee09,uzdensky10,loureiro12,huang13},
whose reconnection rate is higher than in the SP reconnection. 
In these studies, the plasma beta is usually set to high.
However, the plasma beta is extremely low in possible reconnection sites
such as the solar corona.\citep{gary01}
In low-beta plasmas, the number of the research on the plasmoid-dominated reconnection
is quite limited,\citep{ni12,baty14} and it is totally unclear
how the above low-beta effects alter the overall evolution. 
It is of importance to understand
plasmoid physics in low-beta plasmas as a basic element,
in order to explore complex systems such as the plasmoid-dominated reconnection.
We also note that recent researches suggest that
the plasmoid-dominated reconnection contains Petschek-like mini-reconnection.\citep{baty12,nakabo13}
In such a case, the above study is
of direct relevance to understand the system.

The purpose of this paper is to extend our recent work (Paper I)
with a better MHD solver
and
to investigate the plasmoid structure in low-beta plasmas in unprecedented detail.
We discuss several advanced structures in the plasmoid system,
such as triple-shock intersections and shock-diamonds.

\section{Numerical setup}
\label{sec:setup}
We use a Godunov-type code to solve resistive MHD equations.
The code is based on one in Paper I and
contains several improvements since then.
It employs the HLLD solver.\citep{miyoshi05}
Compared with the HLL solver in Paper I,
the HLLD solver additionally resolves
contact and rotational discontinuities inside the Riemann fan.
The surface values are interpolated by the MUSCL scheme.
The minmod limiter is applied to primitive variables.

The initial setup is similar to the model in Paper I.
The simulations are carried out in the $x$-$z$ two-dimensional plane.
We set the reconnection point to $(x,z) = (0,0)$.
We solve one quadrant of $[0,200] \times [0,150]$
with 12000 $\times$ 9000 grid points.
The resolution is twice better than in Paper I.
We use a symmetric boundary at $x=200$ and a reflecting wall at $z=150$.
We consider a Harris sheet:
$\vec{B}(z) = \tanh(z) \vec{\hat{x}}$, $\vec{v}=0$,
$\rho(z) = 1 + \cosh^{-2}(z)/ \beta_{\rm up}$,
$p(z) = 0.5 ( \beta_{\rm up} + \cosh^{-2}(z) )$,
where $\beta_{\rm up}$ is the plasma beta in the upstream region.
These symbols have their standard meanings, unless stated otherwise.
The Alfv\'{e}n velocity in the upstream region is set to $c_{\rm A,up} = |B|/\sqrt{\rho}=1$.
Thus the unit time corresponds to the Alfv\'{e}n transit time across the Harris sheet. 
We set the polytropic index to $\gamma = 5 / 3$.

We employ a localized resistivity,
$\eta(x,z) = \eta_0 + (\eta_1-\eta_0)\cosh^{-2} ( \sqrt{x^2+z^2} )$,
where $\eta_1=1/60$ and $\eta_0=1/1000$. 
A weak perturbation $\delta A_y = -0.06 \exp[-(x^2+z^2)/4]$ is imposed
to magnetic fields to help the reconnection onset.
In this way, we invoke Petschek-type reconnection\citep{petschek}
to obtain a single, giant plasmoid at the jet front.
We intend to understand basic structures
in this idealized configuration.
We study two representative cases: $\beta_{\rm up}=0.1$ in Run 1 and $\beta_{\rm up}=0.2$ in Run 2.

\section{Results}
\begin{figure}
\centering
\includegraphics[width=\columnwidth]{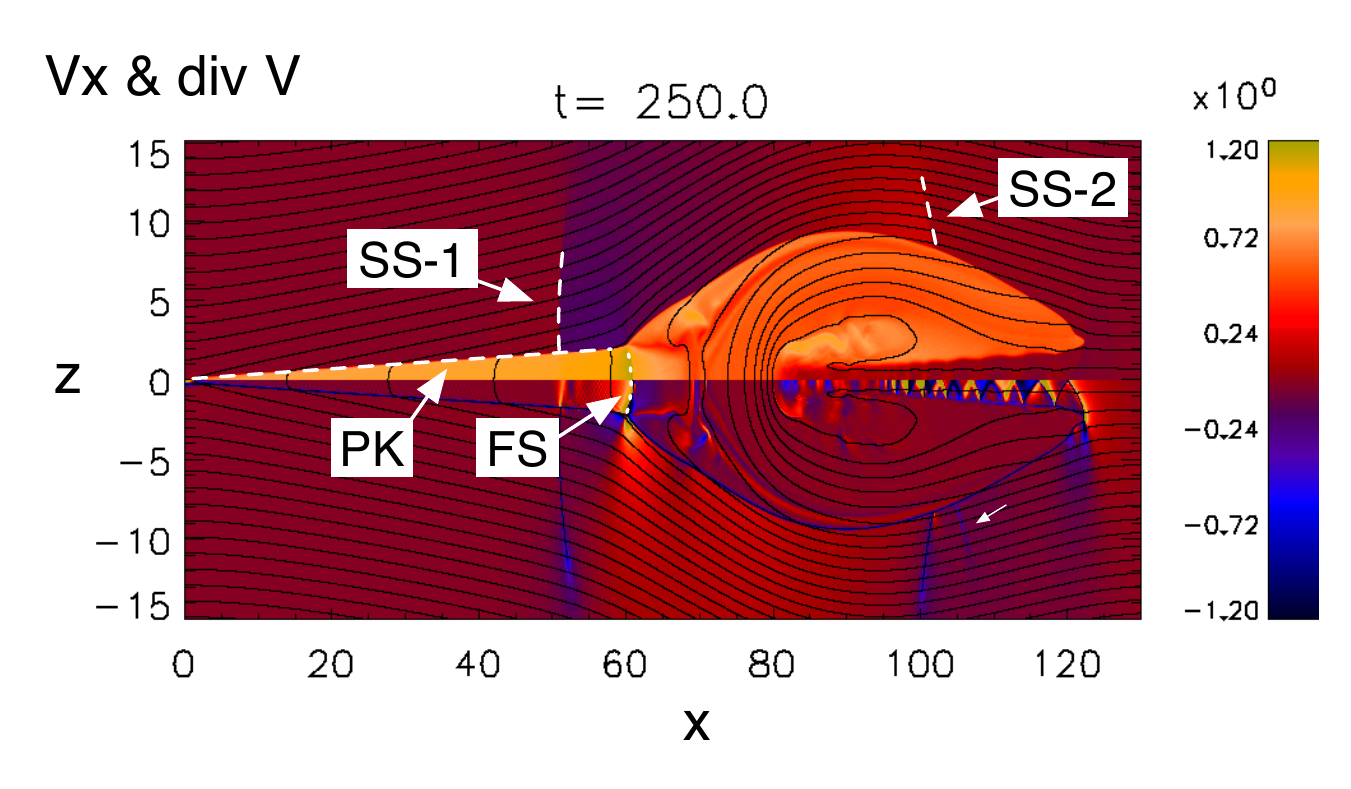}
\caption{
\label{fig:snapshot}
(Color online)
Snapshots at $t=250$ in Run 2, overlaid by magnetic field lines.
(Top) Plasma outflow velocity $v_x$.
(Bottom) The divergence of the velocity field ${\nabla \cdot} \vec{v}$, magnified by 5.
}
\end{figure}

We examine Run 2, which corresponds to the main run in Paper I. 
Figure \ref{fig:snapshot} is a snapshot at $t=250$.
The upper half shows the outflow velocity $v_x$ and
the lower half shows the divergence of the velocity field ${\nabla \cdot} \vec{v}$.
As discussed in Paper I, there are characteristic shocks, such as
a Petschek slow shock (PK),
a reverse fast shock (FS),
a postplasmoid vertical slow shock (SS-1),
and a forward vertical slow shock (SS-2).
In the lower half, these shocks are evident in
compressional layers of ${\nabla \cdot} \vec{v}<0$. 
The postplasmoid vertical slow shock is sometimes found
on the backside of the plasmoid. 
Such configuration is similar to a normal shock on the airfoil,
when an aircraft is flying at a transonic speed.\citep{anderson02}
These shocks are features of low-beta reconnection,
in which an Alfv\'{e}nic outflow drives the plasmoid
at a transonic or supersonic speed.
%sometimes called recompression shock,
%the air (the ambient plasma) is first compressed
%on the front side of the airfoil (plasmoid) and then compressed again by the normal shock.

\begin{figure}
\centering
\includegraphics[width=\columnwidth]{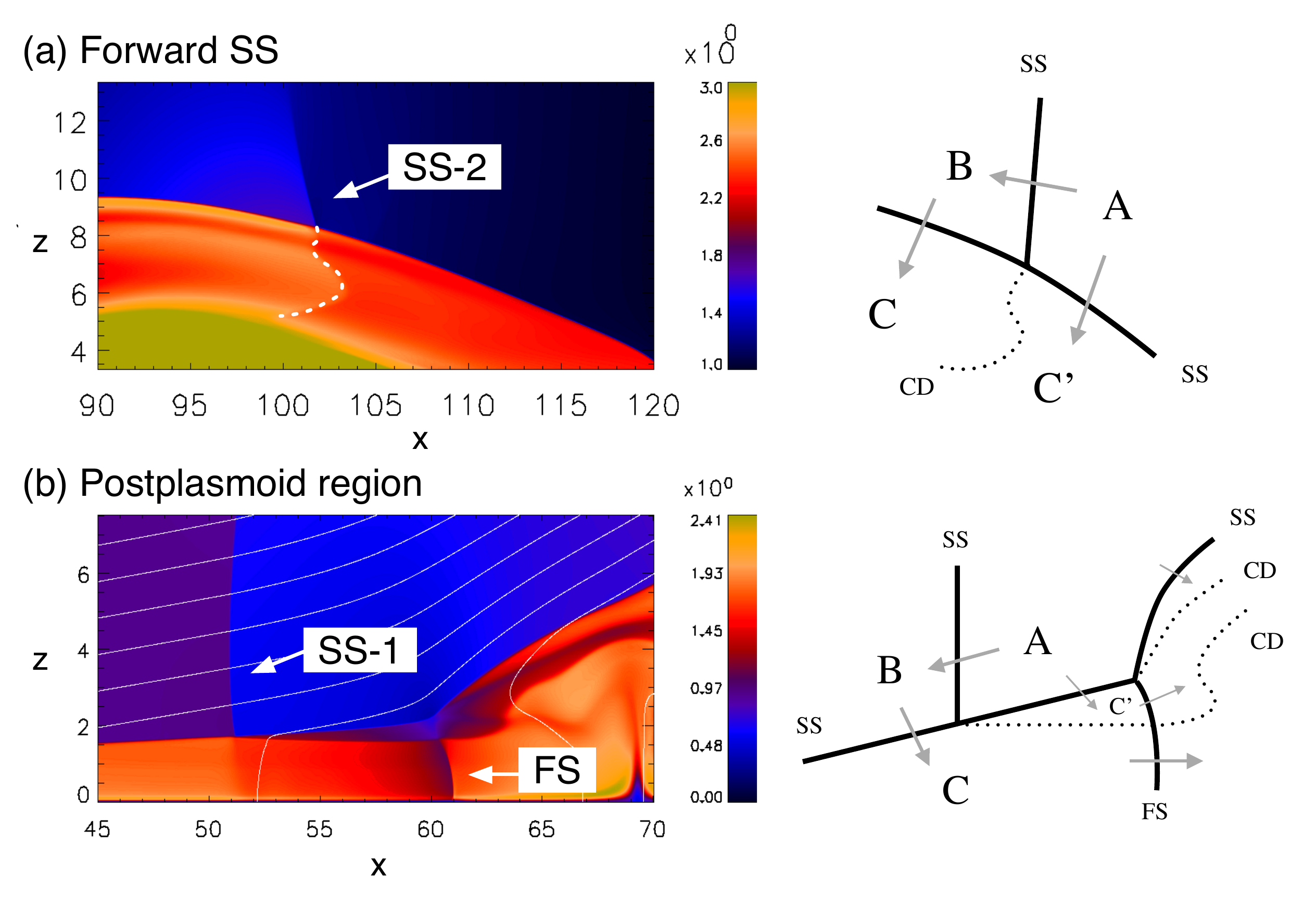}
\caption{
\label{fig:SS}
(Color online)
Plasma density $\rho$ and schematic drawings in
(a) the forward slow-shock region, and
(b) the postplasmoid region.
}
\end{figure}

%\subsection{Triple-shock intersection}
The forward slow-shock connects with the plasmoid boundary.
Since the plasmoid boundary is a slow shock,\citep{ugai95}
it is interesting to see how three shocks intersect each other.
Shown in Figure \ref{fig:SS}(a) is the plasma density
near the forward slow-shock.
The dense (orange) region corresponds to the plasmoid.
The forward slow-shock meets
the plasmoid boundary at $(x,z) \approx (102,8.5)$.

In hydrodynamics, we expect another shock or discontinuity
from such a triple-shock intersection.\citep{landau59}
Consider the fluid advection from the upstream $A$ region to the downstream plasmoid.
There is one shock between in $A \rightarrow C'$,
while there are two in $A \rightarrow C$.
Since the Rankine-Hugoniot relation is nonlinear,
the fluid states should be different in $C$ and $C'$.
Then a shock or a discontinuity should
separate the two regions. 
We expect that this argument holds true in MHD,
because MHD is a complex extension of hydrodynamics.
In this case, there is a contact discontinuity (CD),
as indicated by the white dashed line in Figure \ref{fig:SS}(a).
The specific entropy ($p/\rho^\gamma$; not shown) is useful to to visualize it.

%We do not recognize such discontinuity for the preshock,
%probably because it is too weak.

Shown in Figure~\ref{fig:SS}(b) is
the plasma density in the outflow region behind the plasmoid.
The fast shock is located at $x \approx 61$.
The plasma density decreases in the upstream side of the shock,
because an adiabatic-type acceleration takes place there.\citep{shimizu00}
As illustrated, there are two triple-shock intersections.
We similarly expect additional shocks or discontinuities.
A contact discontinuity stems from the left intersection point,
where the vertical slow-shock connects with the Petschek shock.
This CD was suggested, but unclear in Paper I. 
In this work, the HLLD solver successfully resolves it between the C and C' regions.
The CD survives even in the downstream of the fast shock,
as illustrated by the lower dotted line in the diagram.
In addition, we find another CD
from the right intersection point toward the upper-left direction.
As a result, between the two discontinuities,
a narrow low-density channel is formed inside the plasmoid. 
The divergence ${\nabla \cdot} \vec{v}$ does not mark them
in the lower half of Figure \ref{fig:snapshot}.
This suggests that these structures are discontinuities rather than shocks.
These discontinuities inside the plasmoid have never been discussed in previous works.

We also find a very weak compressional layer at $x=105$,
as indicated by the small arrow in Figure \ref{fig:snapshot}.
It sits in the front side of the normal shock
since the earlier stage of the simulation.
Unfortunately
the layer is too faint to obtain shock quantities.
For example, our Rankine-Hugoniot analysis tells us
that slow-mode Mach numbers are around $2$
in both upstream (right) and downstream (left) regions,
but it is difficult to evaluate the shock-propagating speed.
At this point, we are not sure whether
it is a compressional layer that could evolve into a slow shock,
or the HLLD solver does not adequately resolve a slow shock.

%\subsection{Shock diamonds}

Next, we discuss another key player, {\it shock-diamonds}.
The plasmoid exhibits a crab-claw structure,
because a dense plasma sheet lies at the center. 
When the bifurcated flows are faster than
the sound speed inside the plasma sheet,
the twin edges invoke oblique shocks,
resulting in a diamond-pattern inside the plasma sheet.\citep{zeni10b,zeni11a}
Such shock-diamonds are evident
in the right side in Figure \ref{fig:snapshot}
(see also Fig.~5b in Paper I).

\begin{figure}
\centering
\includegraphics[width=\columnwidth]{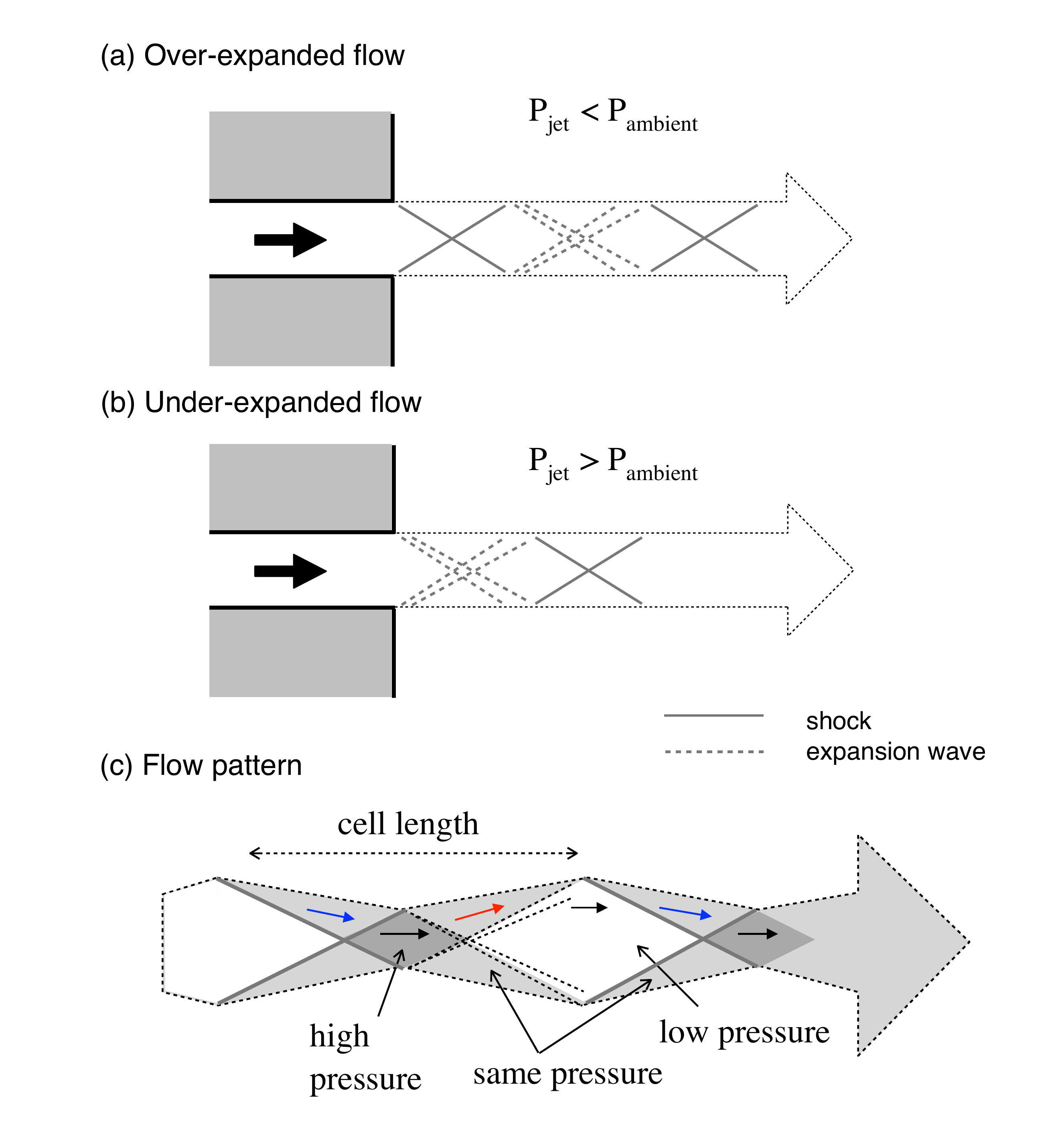}
\caption{
\label{fig:nozzle}
(Color online)
Supersonic nozzle problem.
(a) Over-expanded flow,
(b) under-expanded flow, and
(c) the flow pattern in the shock cell.}
\end{figure}

In the rest frame of the plasmoid,
a narrow plasma sheet comes in at a supersonic speed.
This is equivalent
to a supersonic nozzle problem in aerospace engineering.\citep{shapiro53,anderson02}
Let us consider that a supersonic flow emanates from a nozzle,
as illustrated in Figure \ref{fig:nozzle}. 
It is classified to over-expanded and under-expanded cases. 
If the jet pressure is lower than the ambient pressure,
the flow is said to be over-expanded.
In this case, the oblique shock propagates inward from the nozzle. 
The jet boundary reflects the oblique shock as an expansion wave,
and then reflects the expansion wave as a compressional (shock) wave.
This cycle results in a diamond-shaped pattern in the downstream. 
When the jet is under-expanded,
i.e., at higher pressure than the ambient pressure,
an expansion wave appears first, and then the same cycle continues. 
The bottom panel in Figure \ref{fig:nozzle} shows the flow pattern in this cycle. 
The flow converges in the downstream of the oblique shock,
and diverges after crossing the high-pressure cell.
Due to the pressure difference,
the jet boundary also exhibits cyclic expansion/compression. 
%This is evident in our simulation, if we take a closer look at the shock-diamond boundary. 
The sign of the vertical velocity changes accordingly,
as indicated by blue and red arrows.

Next we examine the role of the shock-diamonds.
Figure \ref{fig:1D} presents 1D cuts of selected quantities
at the midplane, $z=0$.
They exhibits periodic oscillation in the shock-diamond region
of $95 \lesssim x \lesssim 120$.
The magenta curve in Figure \ref{fig:1D} indicates
the sound speed toward the $+x$ direction.
Note that 
the sound speed is virtually identical to the fast-mode speed
in the right side of the tangential discontinuity (TD) at $x \approx 80$,
which separates the reconnected magnetic field and the initial plasma sheet.
It is $c_s \approx 0.42$ in the rightmost plasma sheet.
This is also indicated by the dashed line in magenta for comparison.
The propagation speed of the TD is $v_{\rm TD} \approx 0.54$,
similar to the plasma speed around the TD.
Since it exceeds $0.42$, macroscopically,
the plasmoid interacts with the plasma sheet at a supersonic speed.
In the plateau of $85 \lesssim x \lesssim 95$
the velocity is $v_x \approx 0.1$ and so
the right-running sound speed is $(v_x+c_s) \approx 0.58 > v_{\rm TD}$.
Between the plateau and the leftmost plasma-sheet,
we recognize gradual slopes in flow properties,
as indicated by the dashed lines.
These results suggest that
the shock-train decelerates the plasma-sheet flow
from a supersonic speed to a subsonic speed
in the comoving frame of the plasmoid. 

In fluid dynamics, it is known that
such a shock-train decelerates a supersonic flow.\citep{shapiro53,matsuo99}
Similar examples include a supersonic Fanno flow and an oblique shock diffuser.
In Figure \ref{fig:snapshot},
the central shock channel becomes narrower and
the Mach angle of each shocklet becomes steeper,
as the plasma sheet penetrates deeper into the plasmoid.
The typical length of the shock-train region is
$10$--$20$ times long as the flow width.\citep{matsuo99}
These are common features of the shock-train that decelerates a supersonic flow. 
From the right upstream to the left downstream in Figure \ref{fig:1D},
the pressure increases in the shock-diamond region.
This is mainly due to adiabatic heating.
The strongest shock in the shock-train is the rightmost oblique shock,
but its Mach number is only $\mathcal{M}\approx 1.2$.
The entropy gain is limited across such a weak hydrodynamic shock.
It appears that
the shock-train works as a quasi-adiabatic decelerator.

We expect shock-diamonds,
wherever a narrow jet crosses a boundary at a supersonic speed.
Motivated by this, we find another shock-diamonds.
Figure \ref{fig:b01}(a) shows the vertical velocity $v_z$ near the intersection of
the Petschek shock and the normal shock at $t=275$ in Run 1
(see also Fig.~6(a) in Paper I).
Note that the range is carefully set to $-0.02 < v_z < 0.02$
in the unit of $c_{A,up}$. 
The normal shock is indicated by the white dash line.
As evident in the triangle pattern in $v_z$,
one can recognize new shock-diamonds inside the reconnection jet.
They are of under-expanded type, because
the pressure in the B region is higher than in the A region, i.e., $p_b > p_a$.
As color-coded in Figure \ref{fig:b01}(a) and
as illustrated in Figure \ref{fig:b01}(a'),
the upward motion in the expansion phase comes first in Figure \ref{fig:b01}
(see also Fig.~\ref{fig:nozzle}c).
The vertical speed is small, $|v_z| \approx 0.01$, and
the fluctuation in pressure is $\Delta p/p < 0.1$.
The shock-diamonds are marginally visible in Run 2,
but they are much more apparent in Run 1.
They further evolve longer in larger simulations. 

Let us examine the condition for the new shock-diamonds.
The shock-diamonds will appear
when the reconnection jet supersonically outruns the normal shock,
$v_{\rm jet} \approx c_{\rm A,up} > c_{\rm s,jet} + v_{\rm shock}$,
where $c_A$ is the Alfv\'{e}n speed and $c_s$ is the sound speed
and the subscripts ``up,'' ``jet,'' and ``shock'' denote
the upstream region, the outflow jet, and the normal shock, respectively. 
We note that plasmas in the B region are stationary and
its properties are similar to the initial upstream plasmas.
Since the normal shock is a slow shock in a low-beta plasma,
we expect that the shock-propagating speed is subsonic, $v_{\rm shock} < c_{\rm s,up}$.
Then the sufficient condition for shock-diamonds is
\begin{equation}
c_{\rm A,up}
>
\sqrt{ \frac{\gamma p_{\rm jet}}{\rho_{\rm jet}} }
+
\sqrt{ \frac{\gamma p_{\rm up}}{\rho_{\rm up}} }
.
\label{eq:threshold}
\end{equation}
We further utilize the pressure balance across the Petschek shock,
\begin{equation}
%p_i\frac{1+\beta}{\beta} <
p_{\rm jet}
%= p_{up}\frac{1+\beta}{\beta} + \rho_{up} v^2_{up}
\approx p_{\rm up}\frac{1+\beta_{\rm up}}{\beta_{\rm up}} + \rho_{\rm up} (0.1 c_{\rm A,up})^2
,
\end{equation}
where the last term is negligible,
$\sim (p_{\rm up}/\beta_{\rm up})\mathcal{O}(10^{-2})$.
The compression ratio is known to be\citep{soward82,birn11}
\begin{equation}
\frac{\rho_{\rm jet}}{\rho_{\rm up}} = \frac{\Gamma(1+\beta_{\rm up})}{1+\Gamma\beta_{\rm up}}
,
\end{equation}
where $\Gamma= {\gamma}/({\gamma-1})$.
Using these relations, Equation \eqref{eq:threshold} yields
%\begin{equation}
%1
%+
%\sqrt{ \frac{1+\beta}{\beta} }
%\sqrt{ \frac{1+\Gamma\beta}{\Gamma(1+\beta)} }
%<
%\sqrt{ \frac{2}{\gamma\beta} }
%\end{equation}
\begin{equation}
\sqrt{ \Gamma\beta_{\rm up} }
+
\sqrt{ 1+\Gamma\beta_{\rm up} }
<
\sqrt{ \frac{2\Gamma}{\gamma} }
.
\end{equation}
This gives a simple condition,
\begin{equation}
\beta_{\rm up} < 0.135.
\end{equation}
Run 1, $\beta_{\rm up}=0.1$, is in this regime.
In the $\beta_{\rm up}\ll 1$ limit,
shock-diamonds will always appear inside the Petscheck outflow.

%One can also see that
%the entire outflow gradually becomes faster in the right direction.
%This is due to the adiabatic acceleration
%near the fast shock (FS in Fig.~\ref{fig:snapshot}) \citep{shimizu00}.

Surprisingly,
the post-plasmoid structure is more than
a superposition of the contact discontinuity (CD) and the shock-diamonds.
Figures \ref{fig:b01}(b) and (c) show
the plasma outflow $v_x$ and
the specific entropy $s=p/\rho^\gamma$. 
Our interpretation is illustrated in Figure \ref{fig:b01}(b').
The entropy, which is an excellent measure of the fluid state,
visualize the CD
from the triple-shock intersection in the horizontal direction.
This is indicated by the white dotted line.
Interestingly, the field lines are bent below the CD.
The bent point propagates inward, as indicated in the dot-dash line.
We think that this is a slow expansion wave (SE).
The $C''$ region between the CD and the wavefront
corresponds to the slow expansion fan.
The following results further support our interpretation.
The narrow angle of the $C''$ region tells us that
the wavefront propagates at the local Alfv\'{e}n speed or slow-mode speed. 
Possible candidates are slow shock, slow expansion wave, and rotational discontinuity. 
From the upstream ($C'$) to the downstream ($C''$), 
the magnetic tilt angle changes from ${\approx}90^\circ$ to ${\approx}50^\circ$.
Since the wavefront angle is only about $-5^\circ$,
the rotational discontinuity is ruled out.
Since both the tangential magnetic field and
the magnetic field strength increase across the wavefront,
the slow shock is ruled out. 
Therefore we conclude that this is a switch-on type slow expansion wave. 
Since plasma beta is high in the outflow exhaust, 
changes in plasma properties are rather subtle to distinguish.
The entropy remains unchanged (Fig.~\ref{fig:b01}(c)) in $C$ and in $C''$,
as it should be.

The formation mechanism of the expansion fan can be understood as follows.
In the upstream side of the Petschek shock,
Alfv\'{e}n speeds are $c_{A}=1.05$--$1.08$ in the rarefied A region
and $c_{A}=0.9$ in the shocked B region.
Accordingly, from the Petschek balance condition,
we expect that the outflow speed in C' is slightly faster than in C. 
These flows are separated by the CD.
In hydrodynamics, the tangential velocities can be different across the CD.
This is why the CD is often called the ``slip line''
in the context of the shock theory.\citep{bendor07}
However, in this MHD case, the two flows are threaded by
the same magnetic field lines inside the outflow region. 
So, as evident in $v_x$ in Fig.~\ref{fig:b01}(b),
the fast flow in C' drags plasmas in C'' via the magnetic field line.
This explains why the field lines are bent at the wavefront
and why the C'' region is slightly rarefied.

\section{Discussion}
\label{sec:discussion}

Thanks to the HLLD solver and the very high resolution,
we have identified contact discontinuities
near the triple-shock intersection points.
We have also examined two shock-diamonds.
One is inside the plasmoid and
works as an adiabatic decelerator for the incoming plasma-sheet flow.
The other is newly found inside the Petschek outflow. 
The post-plasmoid structure further involves the slow expansion fan. 
Many of them arises from intersections between
the classical Petschek slow shocks and the normal shocks,
the latter of which appear in the $\beta<1$ regime (Paper I). 
In MHD, the discontinuities and slow-mode structures develop and persists,
unless they are blurred by the nonideal effects. 
At this stage of investigation,
their impact on the overall evolution is not clear,
however, the system certainly becomes more complex than expected. 
They were not adequately resolved by the HLL solver, but
the HLLD solver successfully resolves them. 

We believe that these features are ubiquitous in low-beta reconnection,
as long as the MHD approximation holds true.
While this work has focused on Petschek reconnection,
the plasmoid interacts with the plasma sheet either
in Sweet--Parker and plasmoid-dominated reconnections as well.
Once a plasmoid moves at a sufficiently high speed,
it is followed by normal shocks
in whatever form of reconnection
(e.g., Fig.~3 in Ref.~\onlinecite{zeni10b}). 
Indeed, in our preliminary results on plasmoid-dominated reconnection,
we confirm that plasmoids invokes a lot of normal shocks. 
This makes the upstream region highly dominated with shocks.
Furthermore, recent researches report that
the plasmoid-dominated reconnection involves
Petschek-type structure with a pair of slow shocks.
\citep{baty12,nakabo13}
Finally, during the turbulent regime,
a plasmoid not only hits the plasma sheet ahead of it,
but also collides into another plasmoid. 
All these considerations suggest that
shocks and discontinuities further intersect with each other
in the plasmoid-dominated regime. 

We are aware of some limitations in our model.
From the application viewpoint,
we have to consider several non-MHD effects. 
For example,
the CD will be blurred by collisionless mixing along the field lines
in the magnetotail.
In solar corona,
the heat conduction along the field lines, radiative cooling,
and collision with the neutral atoms will modify the scenario. 
Our results will be a baseline to discuss these non-MHD effects. 
%The necessary condition for the fast shock is $\beta<1$.
%We expect that these shock-diamonds are generic
%in low-beta reconnection of $\beta_{\rm up}\ll 1$. 
%They may lead to non-uniform plasma emission.
From the numerical viewpoint,
the HLLD solver neglects slow waves inside the Riemann fan.
Therefore, it basically tends to blur slow-mode structures. 
In particular, in a low-beta plasma,
the fast-mode and Alvf\'{e}n speeds are much higher than the slow-mode speed
in the field-aligned direction.
This is one of the reasons why the compressional layer is ambiguous in Section 3.
The slow expansion fan is obviously beyond the scope of the HLLD solver.
To resolve these slow-shock structures,
we may need to employ an exact Riemann solver.\citep{takahashi14}

\begin{figure}
\centering
\includegraphics[width=\columnwidth]{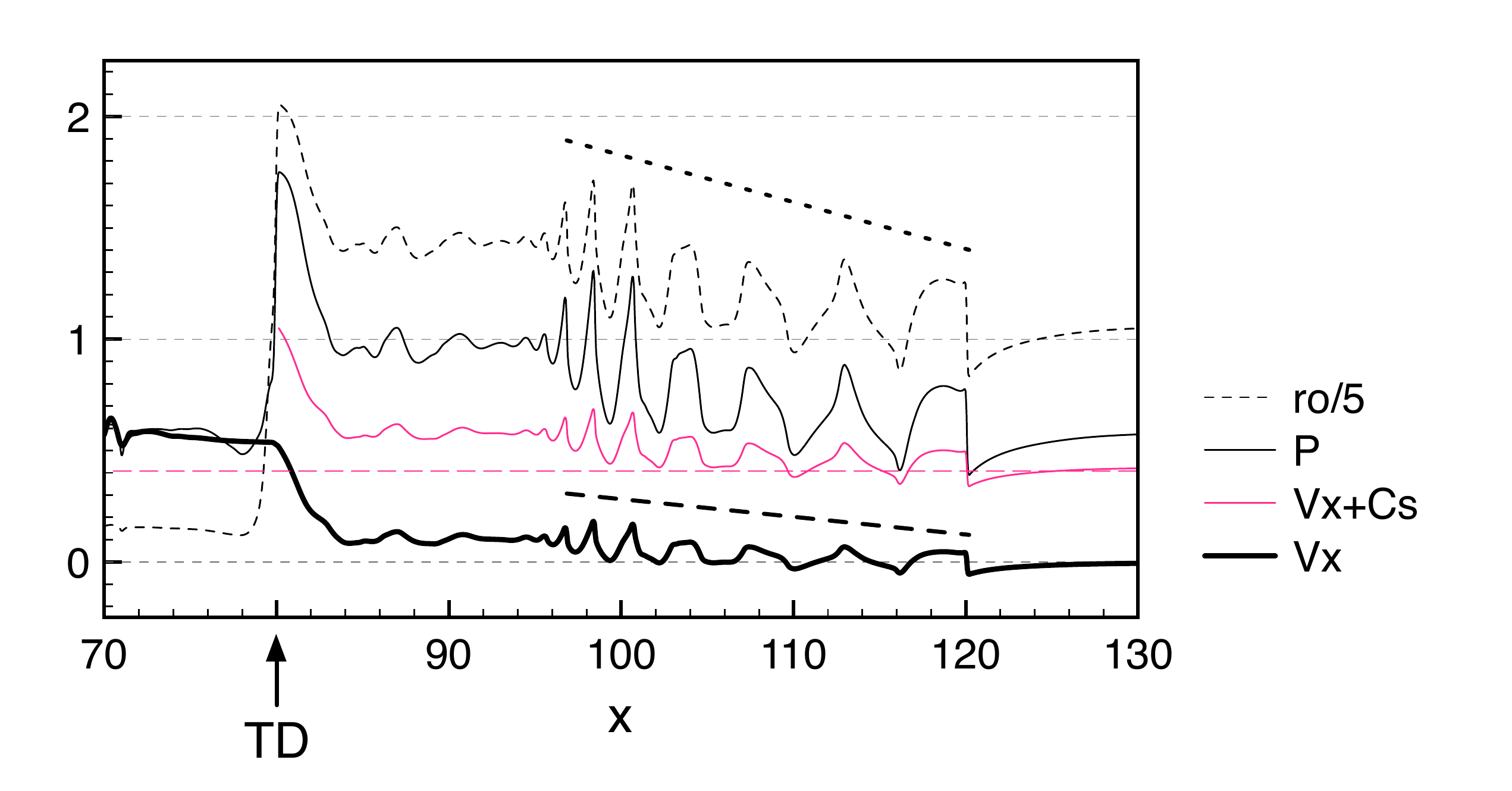}
\caption{
\label{fig:1D}
(Color online)
1D cuts of MHD quantities at $z=0$.
The plasma density ($\rho$), the pressure ($p$),
the right-running sound speed ($v_x+c_s$; only at $x>80$),
and
the plasma speed ($v_x$) are preseted.
The `TD' indicates the position of
the tangential discontinuity.
}
\end{figure}

\begin{figure}
\centering
\includegraphics[width=\columnwidth]{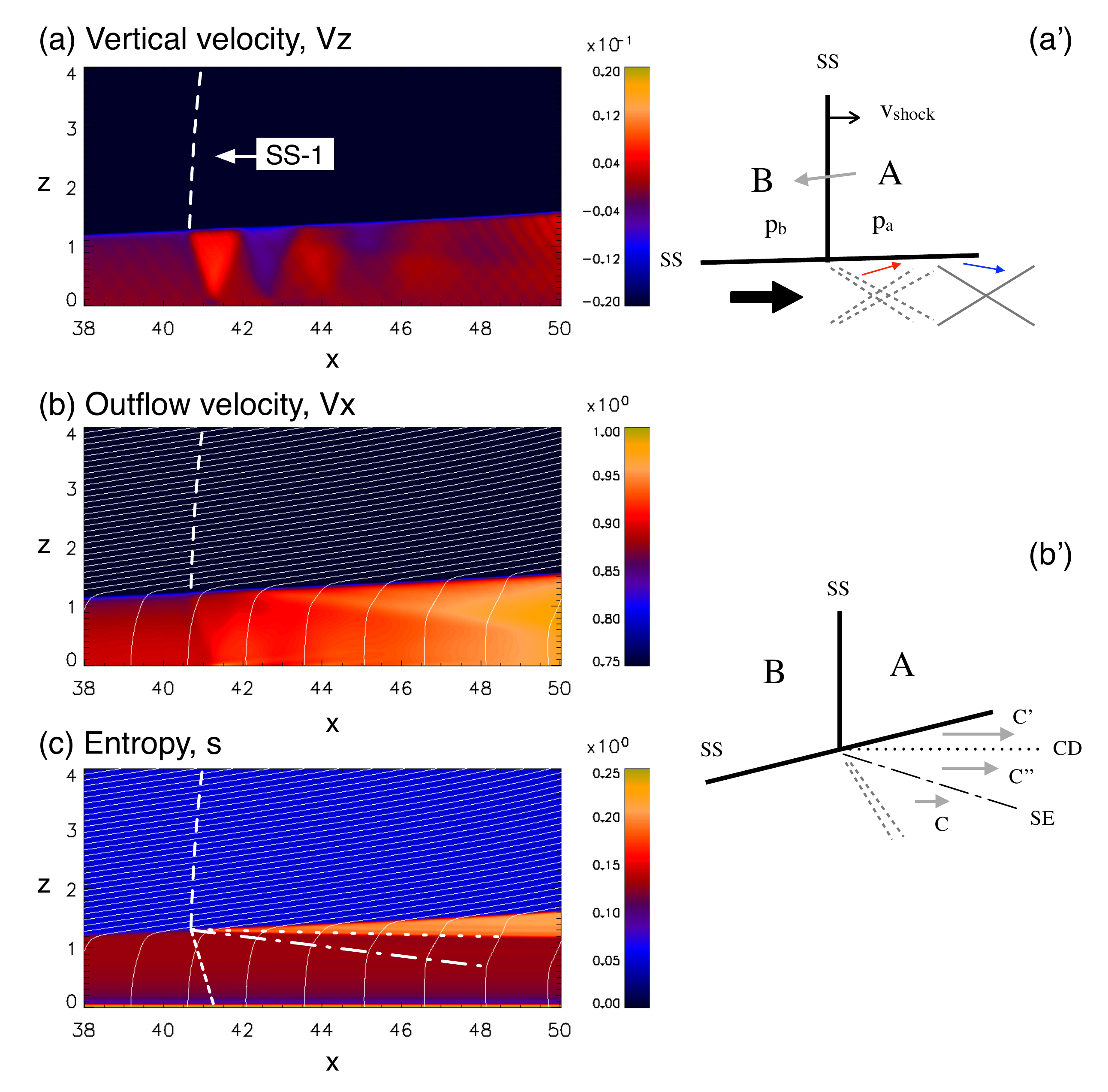}
\caption{
\label{fig:b01}
(Color online)
Properties near the shock intersection point in Run 1,
(a) vertical velocity $v_z$,
(b) outflow velocity $v_x$, and
(c) specific entropy $s=p/\rho^\gamma$.
The vertical dash lines indicate the normal shocks.
The bottom two panels are overlaid by the magnetic field lines. 
In (a) and (b), the ranges are carefully adjusted to emphasize the structure.
The right panels schematically illustrate
(a') shock-diamonds and (b') discontinuities.
}
\end{figure}

\begin{figure}
\centering
\includegraphics[width=\columnwidth]{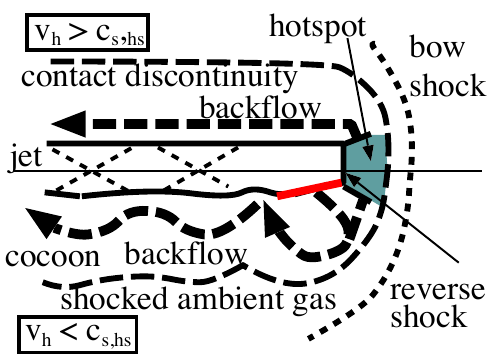}
\caption{
\label{fig:mizuta}
(Color online)
Structure of the astrophysical jet.
(Reproduced from \citet{mizuta10}; Courtesy of A. Mizuta)
}
\end{figure}

It is interesting to mention that
our single plasmoid system is somewhat similar to the astrophysical jet system.
Figure \ref{fig:mizuta} schematically illustrates
a typical structure of the jet.\citep{mizuta10}
Aside from the minor difference between the upper and lower panels,
we outline the essence in the following way.
A narrowly-collimated jet travels at a supersonic speed,
hits the interstellar matter, form a reverse shock in front of the CD,
and then turns round as a backflow inside the cocoon.
The supersonic jet is thought to contain oblique shocks,
which may be observed as periodic knots.\citep{godfrey12}
The oblique shocks are attributed to either 
%compressional shocks %(referred to as recollimation shocks or reconfinement shocks)
shock-diamonds or similar periodic structure from the jet-launching site
\citep{norman82,sanders83,daly88,kom97,matsumoto12}
or shocks driven by the jet-cocoon interaction. 
In our system,
in the comoving frame of the plasmoid,
the narrow plasma-sheet travels at a supersonic speed,
hits a dense plasma region in front of the tangential discontinuity,
and then turns round as a backflow inside the plasmoid. 
%The backflow further drives the Kelvin-Helmholtz instability,
%as shown in Section III E in Paper I. 
Interestingly, we do not recognize a reverse shock inside the plasmoid.
Plasmas are compressed around $80<x<84$ in Figure \ref{fig:1D}, but
there is a gradual slope in front to the TD.
This is because the flow is already decelerated by the shock-train.
The reverse shock does not stand,
because the TD propagation speed is slower than
the the right-running sound speed, $(v_x+c_s) \approx 0.58 > v_{\rm TD}$. 
We attribute this to over-expanded shock-diamonds. 
As the ambient medium confines the jet,
a supersonic flow tends to be adiabatically decelerated in this case.
This will also enhance the interaction between the jet and the surrounding medium.
In contrast, the astrophysical jet is usually under-expanded.\citep{norman82}
Since released to a tenuous space,
a supersonic jet can be adiabatically accelerated.
It is reasonable that the deceleration effect is less prominent.
In summary, the plasmoid structure is similar to the astrophysical jet structure, but
it lacks a reverse shock due to the deceleration effect in overexpanded shock-diamonds.

In this work, we have explored advanced shock-structures
around the plasmoid in magnetic reconnection in low-beta plasmas,
such as triple-shock intersections and two shock-diamonds.
The new structures will be basic elements to
discuss various reconnection-plasmoid systems in low-beta plasmas.
From the viewpoint of basic fluid physics,
these structures are the outcome of
compressible fluid effects or high-speed fluid dynamics.\citep{shapiro53,anderson02}
An adiabatic effect \citep{shimizu00} and a new class of shocks (Paper I)
also belong to compressible effects.
These compressible effects appear,
when the typical flow speed is comparable with or is faster than the sound speed. 
Low-beta plasmas usually satisfy this condition,
because the Alfv\'{e}nic motion easily becomes transonic or supersonic. 
The compressible fluid dynamics will be a key to understand
magnetic reconnection and associated MHD phenomena in low-beta plasmas,
including the plasmoid-dominated reconnection.

\begin{acknowledgments}
This work was supported by Grant-in-Aid for Young Scientists (B) (Grant No. 25871054).
Simulations were carried out at Center for Computational Astrophysics,
National Astronomical Observatory of Japan. 
The HLLD MHD code is available from the author upon request. 
\end{acknowledgments}

\clearpage

\end{document}